\documentclass[onecolumn]{template}   

\usepackage{graphicx}         
\usepackage{amsmath}
\usepackage{amsfonts}
\usepackage{enumerate}   
\usepackage{bm}   
\usepackage{amssymb}
\usepackage{chngcntr}
\usepackage{ dsfont }
\usepackage{color}
\usepackage{hyperref}

\newcommand{\Y}{\sigma}
\newcommand{\dd}{\mbox{{\normalfont d}}}
\newcommand{\sgn}[2]{\left\lceil#1\right\rfloor^{#2}}

\begin{document}

\begin{frontmatter}

\title{EDCHO: High Order Exact Dynamic Consensus}

\thanks[footnoteinfo]{\textcolor{red}{This is the preprint version of the accepted Manuscript: Rodrigo Aldana-López, Rosario Aragüés, Carlos Sagüés,
“EDCHO: High order exact dynamic consensus”, Automatica, Volume 131, 2021, ISSN 0005-1098. DOI: 10.1016/j.automatica.2021.109750.
Please cite the publisher's version. For the publisher's version and full citation details see:
\url{https://doi.org/10.1016/j.automatica.2021.109750.}\\
Simulation files for the algorithms presented in this work can be found on \url{https://github.com/RodrigoAldana/EDC}
}\\This work was supported by projects COMMANDIA SOE2/P1/F0638
(Interreg Sudoe Programme, ERDF), PGC2018-098719-B-I00 (MCIU/ AEI/
FEDER, UE) and DGA T45-17R (Gobierno de Aragon). The first author
was funded by "Universidad de Zaragoza - Santander Universidades" program.}

\author[First]{Rodrigo Aldana-López*} 
\author[First]{Rosario Aragüés} 
\author[First]{Carlos Sagüés} 

\address[First]{Departamento de Informatica e Ingenieria de Sistemas (DIIS) and Instituto de Investigacion en Ingenieria de Aragon (I3A), 
\\
Universidad de Zaragoza, Zaragoza 50018, Spain.\\
(e-mail: rodrigo.aldana.lopez@gmail.com, raragues@unizar.es, csagues@unizar.es)}

\begin{abstract}               
This article addresses the problem of average consensus in a multi-agent system when the desired consensus quantity is a time varying signal. Although this problem has been addressed in existing literature by linear schemes, only bounded steady-state errors have been achieved. Other approaches have used first order sliding modes to achieve zero steady-state error, but suffer from the chattering effect. In this work, we propose a new  exact dynamic consensus
algorithm which leverages high order sliding modes, in the form of a distributed differentiator to achieve zero steady-state error of the average of time
varying reference signals in a group of agents. Moreover, our proposal is also able to achieve consensus to high order derivatives of the average signal, if desired. An in depth formal study on the stability and convergence for EDCHO is provided for undirected connected graphs. Finally, the effectiveness and advantages of our proposal are shown with concrete simulation scenarios.
\end{abstract}


\end{frontmatter}

\section{Introduction}

In the context of cyber-physical systems, there are a lot of scenarios where the coordination of many subsystems is needed. There is no doubt that distributed solutions are preferred over centralized ones, when big networks of agents are involved in the scenario \cite{Solmaz2017}. This is so, since distributed solutions scale better with respect to the size and topology of the network, and are more robust against failures \cite{Kar2008}. Static consensus, where all subsystems (herein referred as agents) manage to agree on a static value such as the average of certain quantities of interest, is a widely studied topic, see for example \cite{Olfati2007,david2020}. On the other hand, consensus towards a time-varying quantity has recently attracted attention  due to its potential applications such as distributed formation control \cite{Alonso2019}, distributed unconstrained convex optimization \cite{Zhu2012}, distributed state estimation \cite{Aragues2015} and distributed resource allocation \cite{Cherukuri2015} just to give some examples. 

The typical approach, which is widely exposed in \cite{Solmaz2017}, relies in a linear protocol. However, in this case, only practical stability towards consensus can be guaranteed, where the accuracy of the steady state depends on the bounds of the derivative of the reference signals, and  it is improved as the connectivity is increased.
This approach has been studied in the presence of disturbances \cite{Shi2013}, delays \cite{Moradian2019}, switching topologies, \cite{kia2015a}, and event-triggered communication schemes, \cite{Kia2015}. Up to now, the most successful method for the dynamic average consensus is the one discussed in \cite{freeman2019}, which by means of making use of First Order Sliding Modes (FOSM) techniques, manages to achieve exact convergence. A similar approach was used in \cite{Rahili2017} to achieve exact consensus in second order systems and in particular to Euler-Lagrange systems. However, both approaches suffer from the same two disadvantages. First, they consider that the derivative of the reference signals is bounded by a known constant, which may be restrictive in some applications. And the second one is that both use FOSM, which introduce the so-called chattering effect \cite[Chapter 3]{fridman2002}. This effect causes these methods to be dangerous for some control systems and to be sensitive to switching delays and measurement noises.
\subsection{Contributions}
In this work, we propose a new High Order Exact Dynamic Consensus (EDCHO)
algorithm which leverages High Order Sliding Modes (HOSM) in the form of a distributed differentiator to achieve zero steady-state error of the average of time
varying reference signals of a group of agents, and as a consequence alleviating the problem of chattering. In this case, it is only required  that a certain high order derivative of the reference signals differences is known to be bounded by a known constant. Moreover, this method successfully achieves consensus not only to the average of the reference signals, but its derivatives. To the best of our knowledge HOSM techniques hasn't been used in the context of dynamical consensus for this purpose. A preliminary analysis of the EDCHO algorithm was presented in \cite{aldana2020} were only some of its features were shown, particularly by means of simulations. However, different from the previous analysis, here we show a formal proof for convergence of the protocol for arbitrary connected undirected graphs. Concretely, in Section \ref{sec:problem} we present the Exact Dynamic Consensus problem statement. In Section \ref{sec:protocol} we provide the proposed protocol and the stability guarantee in Theorem \ref{th:stability} which is the main result of this work. Moreover, we develop some auxiliary results in Sections \ref{sec:towards}, \ref{sec:contraction} and \ref{sec:parameters} needed to prove Theorem \ref{th:stability}, which are all new with respect to \cite{aldana2020}. Furthermore, a proof of Theorem \ref{th:stability} can be found in Section \ref{sec:convergence}. Finally, we provide simulation examples which corroborate the performance of our proposal in Section \ref{sec:simulations}. 

\subsection{Notation}

Let $\mathbb{R}$ be set of the real numbers and $\mathbb{R}_+=\{x\in\mathbb{R}:x\geq 0\}$. The symbols $\dot{x}(t), \ddot{x}(t)$ represent the first and second time derivatives of $x(t)$ whereas $x^{(\mu)}(t)$ for $\mu\geq 0$ represent the $\mu$-th time derivative of $x(t)$. Let $\mathds{1} = [1,1,\dots,1]^T\in\mathbb{R}^n$ and $I\in\mathbb{R}^{n\times n}$ the identity matrix, where the dimension $n$ is defined depending on the context. Italic indices $i,j$ will be used when referring to agents in a multi-agent system. Let $\text{sign}(x) = 1$ if $x> 0$, $\text{sign}(0)=0$ and $\text{sign}(x)=-1$ if $x<0$. Moreover, if $x\in\mathbb{R}$, let $\lceil x\rfloor^\alpha\triangleq |x|^\alpha\text{sign}(x)$ for $\alpha>0$ and $\lceil x\rfloor^0\triangleq{\text{sign}}(x)$. In the vector case $x=[x_1,\dots,x_n]^T\in\mathbb{R}^n$, then $\sgn{x}{\alpha}\triangleq\left[\sgn{x_1}{\alpha},\dots,\sgn{x_n}{\alpha}\right]^T$ for $\alpha\geq 0$. For any matrix $A\in\mathbb{R}^{m\times n}$, let $s_{\min}(A)$ and $s_{\max}(A)$ represent the smallest and largest singular values of $A$ respectively. For $x\in\mathbb{R}^n$, $\|x\| \triangleq \sqrt{\sum_{i=1}^nx_i^2}$. $\text{diag}(v), v\in\mathbb{R}^n$ represents a diagonal matrix whose diagonal is composed by $v$, and $\text{diag}(A), A\in\mathbb{R}^{n\times n}$ represents a vector composed by the diagonal components of $A$. Moreover, $\text{blockdiag}(\bullet,\dots,\bullet)$ represents the typical block diagonal operator.
 
\section{Problem statement}\label{sec:problem}
The general setting in this work is the following. Consider a multi-agent system consisting of $n$ agents. Each agent $i$ has access to a local time varying signal $u_i(t)\in\mathbb{R}$. Additionally, each agent is capable of communicating with other agents according to a communication topology defined by a connected undirected graph $\mathcal{G}$ (See Appendix \ref{sec:app_graph}). Moreover, each agent $i$ has an output $y_i(t)\in\mathbb{R}^{m+1}$ which, as explained later, corresponds to the variable of interest that must achieve consensus between all agents in the network. Furthermore, each agent $i$ has an internal state $x_i(t) = [x_{i,0}(t),\dots,x_{i,m}(t)]^T\in\mathbb{R}^{m+1}$ which is governed by the dynamic equation
\begin{equation}
    \dot{x}_i(t) = f_i(x_i(t),p_i(t),u_i(t)), \ \ x_i(t_0) = x_{i}^0
    \label{eq:genericF}
\end{equation}
where $p_i\in\mathbb{R}^{d_i}$ is a vector of received messages from its $d_i$ neighbors. The goal of this multi-agent system is stated in the following.

\begin{defn}[Exact Dynamic Consensus] The multi-agent system is said to achieve EDC, if there exists $T>0$ such that the individual output signals for each agent reach
\begin{equation}
\label{eq:prob}
    y_{1,\mu}(t) = y_{2,\mu}(t) = \dots = y_{n,\mu}(t) =  \bar{u}^{(\mu)}(t)
\end{equation}
$\forall t\geq t_0+T$, $\forall\mu\in\{0,\dots,m\}$ and $
\bar{u}(t) = \frac{1}{n}\sum_{i=1}^nu_i(t)$.
\end{defn}

\begin{prob}[High order exact average consensus]
\label{prob:dynamic_consensus}
Given the set of local signals $ \{u_1(t),\dots,u_n(t)\}$, the problem consists in designing the specific protocol of each agent, i.e., choosing the output $y_i(t) = [y_{i,0},\dots,y_{i,m}]^T$ as a function of $x_i$ and $u_i$, which information to share to other agents, and the function $f_i(\bullet,\bullet,\bullet)$ such that the multi-agent system achieve EDC.

\end{prob}

Solutions to Problem \ref{prob:dynamic_consensus} can be used in a variety of applications as described in \cite{Solmaz2017}. Similarly as in \cite{Solmaz2017}, we consider the following assumption.
\begin{assum}
\label{as:initial}
The initial conditions for \eqref{eq:main_algo} are set to be such that $\sum_{i=1}^n x_{i,\mu}(t_0) = 0$, $\forall\mu\in\{0,\dots,m\}$.
\end{assum}
Note that Assumption \ref{as:initial} is trivially satisfied without the need of any global information if all agents set $x_{i,\mu}(t_0)=0, \forall \mu\in\{0,\dots,m\}$.

\section{The EDCHO algorithm}\label{sec:protocol}

The EDCHO algorithm proposed in this work to obtain EDC has the following structure:
\begin{equation}
\label{eq:main_algo}
\begin{array}{rcll}
    \dot{x}_{i,0}(t) &=& x_{i,1}(t) &+k_0\sum_{j=1}^na_{ij}\lceil y_{i,0}(t) - y_{j,0}(t) \rfloor^{\frac{m}{m+1}} \\
    &\ \vdots &&\\
    \dot{x}_{i,\mu}(t) &=& x_{i,\mu+1}(t) &+k_\mu \sum_{j=1}^na_{ij}\lceil y_{i,0}(t) - y_{j,0}(t) \rfloor^{\frac{m-\mu}{m+1}} \\
    &\ \vdots &&\\
    \dot{x}_{i,m}(t) &=& &+k_m \sum_{j=1}^na_{ij}\sgn{y_{i,0}(t) - y_{j,0}(t) }{0} \\
    y_{i,\mu}(t) &=& u_i^{(\mu)}(t)& - x_{i,\mu}(t).
\end{array}
\end{equation}
where each agent has a state $x_i = [x_{i,0},\dots,x_{i,m}]^T$ and $a_{ij}$ are the elements of the adjacency matrix of $\mathcal{G}$. Hence, each agent shares only $y_{i,0}$ to its neighbors, in contrast to sharing all $y_{i,0},\dots,y_{i,m}$ which is not necessary, reducing communication load.  Moreover, the algorithm depends on the the gains $k_0,\dots,k_m>0$ which will be designed as described later in order for \eqref{eq:main_algo} to achieve EDC provided that the following assumption holds.
\begin{assum}
\label{as:bounded}
The signals $u_1(t),\dots,u_n(t)$ all satisfy $\left|\bar{u}^{(m+1)}(t)-u_i^{(m+1)}(t)\right|\leq L, \forall t\geq t_0$ with known $L>0$.
\end{assum}
\begin{rem}
Note that \eqref{eq:main_algo} is a system with discontinuous right hand side. Hence, solutions to \eqref{eq:main_algo} are properly understood in the sense of Filippov \cite{cortes2008}. This is, \eqref{eq:main_algo} is studied as a differential inclusion with $\sgn{0}{0}=[-1,1]$.
\end{rem}
\begin{rem}
Note that if $m=0$, \eqref{eq:main_algo} resembles some of the basic results proposed in \cite{freeman2019}, hence being subsumed by the approach in this work.
\end{rem}
The following is the main result of this work, which states that there exist a non-empty set of possible values for the gains $k_0,\dots,k_m$ such that \eqref{eq:main_algo} achieves EDC.
\begin{thm}
\label{th:stability}
Let Assumptions \ref{as:initial} and \ref{as:bounded}. Moreover, let $k_{\mu}=\lambda_{\mu}k_{\mu-1}^{\frac{m-\mu}{m-(\mu-1)}}$ for $\mu=1,\dots,m$ with $\lambda_1,\dots,\lambda_m$ parameters chosen such that system \eqref{eq:levant_recursive} is finite time stable for $\theta=0$. Therefore, there exists sufficiently large  $k_0>0$ and $T>0$ such that the EDC property is achieved for \eqref{eq:main_algo}.
\end{thm}
The proof of Theorem \ref{th:stability} can be found in Section \ref{sec:convergence} after some needed results which are developed in the following sections. 
\section{Towards convergence of EDCHO}\label{sec:towards}
First, we provide some results which are required to show that \eqref{eq:main_algo} achieves EDC. 
As it will be evident latter, it is convenient to write \eqref{eq:main_algo} with a different set of gains per edge, just as a mere tool for the proof. This is, let $K_\mu = \text{diag}([k_{1,\mu},\dots,k_{\ell,\mu}]), \forall \mu\in\{0,\dots,m\}$ where $\ell$ is the number of edges. Then, the modified version of \eqref{eq:main_algo} is
\begin{equation}
\begin{aligned}
\label{eq:main_algo_vec2}
    \dot{X}_{\mu}(t) &= X_{\mu+1}(t) + DK_\mu\sgn{D^TY_{0}(t)}{\frac{m-\mu}{m+1}}\\ &\text{for } 0\leq\mu \leq m-1, \\
    \dot{X}_{m}(t) &= DK_m\sgn{D^TY_0(t)}{0}\\
    Y_\mu(t) &= U^{(\mu)}(t)-X_\mu(t), \forall \mu\in\{0,\dots,m\}
\end{aligned}
\end{equation}
where $X_\mu=[x_{1,\mu},\dots,x_{n,\mu}]^T, Y_\mu=[y_{1,\mu},\dots,y_{n,\mu}]^T$, $U=[u_1,\dots,u_n]^T$ and $D$ is the incidence matrix of $\mathcal{G}$.  Moreover, Assumption \ref{as:bounded} implies $PU^{(m+1)}(t)\in[-L,L]^n$ with $P = (I-(1/n)\mathds{1}\mathds{1}^T)$. The following is an interesting property of \eqref{eq:main_algo_vec2} which basically states that under Assumption \ref{as:initial}, the trajectories of \eqref{eq:main_algo_vec2} are orthogonal to $\mathds{1}$.
\begin{lem}
\label{le:constant}
Under Assumption \ref{as:initial},
the following identity is satisfied for \eqref{eq:main_algo}:
\begin{equation*}
    \mathds{1}^T X_{\mu}(t) = 0, \ \ \forall t\geq t_0, \ \  \forall\mu\in\{0,\dots,m\}
\end{equation*}
\end{lem}
\begin{pf}
Denote $s_\mu = \mathds{1}^TX_\mu$.
We proceed by induction:
let $\mu=m$ as induction base
$
    \dot{s}_m = \mathds{1}^T\dot{X}_m =-\mathds{1}^T  DK_m\sgn{D^TY_0}{0} = 0
$.
Hence, the value of $s_m(t) = s_m(t_0)=0$ remains constant $\forall t\geq t_0$ under Assumption \ref{as:initial}. Now, assume $s_{\mu+1}(t)=s_{\mu+1}(t_0)=0$,  $\forall\mu\in\{0,\dots,m-1\}$ remains constant  $\forall t\geq t_0$, then,
\begin{equation*}
\begin{aligned}
    \dot{s}_\mu =& \mathds{1}^T\dot{X}_\mu =\mathds{1}^T\left(X_{\mu+1}(t) +  DK_\mu\sgn{D^TY_0}{\frac{m-\mu}{m+1}}\right)\\ =& 
     \mathds{1}^TX_{\mu+1}(t_0) +  \mathds{1}^T DK_\mu\sgn{D^TY_0}{\frac{m-\mu}{m+1}} = 0
\end{aligned}
\end{equation*}
where Assumption \ref{as:initial} was used. Then, $\mathds{1}^TX_\mu(t) = 0, \forall\mu\in\{0,\dots,m\}, \ \forall t\geq t_0$ which concludes the proof.
\end{pf}
It also can be shown that if protocol \eqref{eq:main_algo_vec2} converges, it will converge to a state which complies with the EDC property. To show so, let $\tilde{Y}_\mu(t)=PY_\mu(t)$ with $P=\left(I-(1/n)\mathds{1}\mathds{1}^T\right)$. Then, its dynamics are given by
\begin{equation}
\begin{aligned}
\label{eq:main_algo_error2}
    \dot{\tilde{Y}}_{\mu}(t) &= \tilde{Y}_{\mu+1}(t) - DK_\mu\sgn{D^T\tilde{Y}_{0}(t)}{\frac{m-\mu}{m+1}}\\ &\text{for } 0\leq\mu \leq m-1, \\
    \dot{\tilde{Y}}_{m}(t) &=PU^{(m+1)}(t)- DK_m\sgn{D^T\tilde{Y}_0(t)}{0}
\end{aligned}
\end{equation}
\begin{cor}
\label{th:steady_state}
If there exists $T>0$ such that the state $\tilde{Y}_\mu(t) = 0, \forall\mu\in\{0,\dots,m\}$, $\forall t\geq t_0+T$ is reached, then \eqref{eq:main_algo_vec2} achieves EDC.
\end{cor}
\begin{pf}
If $\tilde{Y}_\mu=0$, then $Y_\mu = (1/n)\mathds{1}\mathds{1}^TY_\mu = (1/n)\mathds{1}\mathds{1}^T(U^{(\mu)} - X_\mu) = \bar{u}^{(\mu)}\mathds{1}$ by Lemma \ref{le:constant}.
\end{pf}

\section{Contraction property of EDCHO}\label{sec:contraction}
In this section we show the so called contraction property as described in \cite{levant2003}. This property states that there exists a non-empty set of gains $K_0, \dots, K_m$ such that trajectories of $\tilde{Y}_\mu(t), \forall\mu\in\{0,\dots,m\}$ gather arbitrary close to the origin in an arbitrary small amount of time. First, we show that it is indeed the case for tree graphs, and then we use this result to show contraction for arbitrary connected graphs.
\subsection{Contraction for tree graphs}
\label{sec:tree}
According to Proposition \ref{prop:vector_decomposition} in Appendix \ref{sec:app_graph},  it is always possible to write $\tilde{Y}_\mu(t)=D\sigma_\mu(t)$ for some $\sigma_\mu(t)\in\mathbb{R}^\ell$ where $\ell$ is the number of edges in $\mathcal{G}$. Now, consider that $\mathcal{G}$ is a tree graph. Note that, in this case, from Proposition \ref{prop:graph}-(\ref{prop:graph_flow}) in Appendix \ref{sec:app_graph},  $s_{\min}(D)>0$ since the flow space of a tree graph has dimension 0 and therefore $D^TD$ is a full rank matrix.
Thus, by writing $U(t) = \bar{u}\mathds{1} + D\tilde{U}(t)$, then under Assumption \ref{as:bounded}, $\tilde{U}^{(m+1)}(t)\in[-\tilde{L},\tilde{L}]^\ell$ with $\tilde{L} = \sqrt{n}L/s_{\min}(D)$ by Proposition \ref{prop:induced_norm} in Appendix \ref{sec:app_ineq}. 

In the following, we study the behaviour of the system, introducing in \eqref{eq:main_algo_error2} the change $\tilde{Y}_\mu = D\sigma_\mu$ to obtain
\begin{equation}
\begin{aligned}
\label{eq:main_algo_vec_error}
    \dot{\sigma}_{\mu}(t) &= \sigma_{\mu+1}(t) -K_\mu \sgn{D^TD\sigma_{0}(t)}{\frac{m-\mu}{m+1}},\\ &\text{\normalfont for } 0\leq\mu \leq m-1 \\
    \dot{\sigma}_{m}(t) &= \tilde{U}^{(m+1)}(t)-K_m \sgn{D^TD\sigma_0(t)}{0} 
\end{aligned}
\end{equation}
By Corollary \ref{th:steady_state}, if $\sigma_\mu=0, \forall\mu\in\{0,\dots,m\}$ is reached, then EDC is achieved. Before showing contraction of \eqref{eq:main_algo_vec_error} we provide some auxiliary results. Write \eqref{eq:main_algo_vec_error} in the recursive form 
\begin{equation}
\label{eq:recursive_2}
    \begin{aligned}
    &\quad\quad\ \ \ \dot{\sigma}_{0}\  = \sigma_{1} -\Lambda_0 \sgn{D^TD\sigma_{0}}{\frac{m}{m+1}}
    \\
    &\mathcal{H}_{m}\left\{\begin{array}{ll}
    \dot{\sigma}_{1}& = \sigma_{2} -\Lambda_{1} \sgn{\sigma_1-\dot{\sigma}_0}{\frac{m-1}{m}}\\
    &\ \vdots\\
    \dot{\sigma}_{\mu} &= \sigma_{\mu+1} -\Lambda_{\mu} \sgn{\sigma_\mu-\dot{\sigma}_{\mu-1}}{\frac{m-\mu}{m-(\mu-1)}}\\
    &\ \vdots\\
    \dot{\Y}_{m} &= -\Lambda_m\sgn{\sigma_m-\dot{\sigma}_{m-1}}{0} + \tilde{U}^{(m+1)}
    \end{array}\right.
    \end{aligned}
\end{equation}
by using the fact that $\sigma_\mu-\dot{\sigma}_{\mu-1} = K_{\mu-1}\sgn{D^TD\sigma_0}{\frac{m-(\mu-1)}{m+1}}$ and defining $K_{\mu}=\Lambda_{\mu}K_{\mu-1}^{\frac{m-\mu}{m-(\mu-1)}}$ for $\mu=1,\dots,m$ and $K_0=\Lambda_0$. This change introduces some advantages because the dynamics in $\mathcal{H}_m$ are decoupled for each component of $\sigma_\mu$, and most importantly the dynamics of each component of $\sigma_\mu$ correspond exactly to the Levant's differentiator error system in recursive form \eqref{eq:levant_recursive}. Hence, by showing that $\dot{\sigma}_0$ resembles the properties of the signal $\theta$ in Appendix \ref{sec:levant}, contraction towards the origin for $\sigma_\mu, \mu\geq 1$ is guaranteed. 
\begin{lem}
\label{lem:integral_bounded}
For any $\delta>0$, $\Omega_0,\dots,\Omega_m>0$ and any trajectory of \eqref{eq:recursive_2} with $\|\sigma_\mu(t_0)\|\leq \Omega_\mu, \forall\mu\in\{0,\dots,m\}$, there exists $K>0$ such that $\int_{t_0}^{t_0+\delta}\|\dot{\sigma}_0(t)\|\dd t<K$.
\end{lem}
\begin{pf}
Choose an arbitrary $\tau>\delta$. Hence, the trajectory of \eqref{eq:recursive_2} for  $t\in[t_0,t_0+\tau]$ satisfying $\|\sigma_\mu(t_0)\|\leq \Omega_\mu$, also satisfy that $\|\sigma_\mu(t)\|\leq \bar{\Omega}_\mu$ in $t\in[t_0,t_0+\tau]$ for some unknown bounds $\bar{\Omega}_\mu$. Moreover, denote with $s_\Lambda = s_{\max}(\Lambda_0)$. Therefore,
\begin{equation*}
\begin{aligned}
&\int_{t_0}^{t_0+\delta}\|\dot{\Y}_0(t)\|\dd t\\
\leq& \int_{t_0}^{t_0+\delta}\ell^{\frac{1}{2m+2}}s_\Lambda\left(s_{\max}(D^TD)\right)^{\frac{m}{m+1}}\left\|{\sigma_0}\right\|^{\frac{m}{m+1}}+\|\sigma_1\|\dd t&\\
\leq& \int_{t_0}^{t_0+\delta}\ell^{\frac{1}{2m+2}}s_\Lambda\left(s_{\max}(D^TD)\right)^{\frac{m}{m+1}}\bar{\Omega}_0^{\frac{m}{m+1}}+\bar{\Omega}_1\dd t \\
=&\left(\ell^{\frac{1}{2m+2}}s_\Lambda\left(s_{\max}(D^TD)\right)^{\frac{m}{m+1}}\bar{\Omega}_0^{\frac{m}{m+1}}+\bar{\Omega}_1\right)\delta = K
\end{aligned}
\end{equation*}
where Corollary \ref{cor:alpha_norm}-(\ref{cor:alpha_norm_lower}) was used with $\alpha=\frac{m}{m+1}$ and Proposition \ref{prop:induced_norm} in Appendix \ref{sec:app_ineq} to introduce $s_\Lambda$ and $s_{\max}(D^TD)$.
\end{pf}
The utility of Lemma \ref{lem:integral_bounded} is that we can use Proposition \ref{prop:levant_boundedness} to fix a desired bound for $\sigma_1$ in the first equation of \eqref{eq:recursive_2} at least for a desired time interval $[t_0,t_0+\delta]$. Then, we can treat $\sigma_1$ as a disturbance with known bound, and focus our attention to designing $\lambda_0\triangleq\min\text{ diag }\Lambda_0$ such that $\sigma_0$ reaches an arbitrarily small vicinity of the origin before that interval ends. This is shown in the following:
\begin{lem}
\label{lem:fo_contraction}
Let $\mathcal{G}$ be a tree and
\begin{equation}
\label{eq:fo_sigma0}
\dot{\sigma}_0(t) = d(t) - \Lambda_0\sgn{D^TD\sigma_0}{\frac{m}{m+1}} 
\end{equation}
 $\sigma_0(t),d(t)\in\mathbb{R}^\ell$, $\overline{d}>0, \Omega_0>0$, and the bounds $\|D\sigma_0(t_0)\|\leq \Omega_0$, $\|d(t)\|\leq \overline{d},  \forall t\in[t_0,+\infty)$. Then, for any $\overline{\delta}>0$ and any $0<\omega_0<\Omega_0$ there exists $0<\overline{\lambda}_0= \min\text{ \normalfont diag } \Lambda_0$ (sufficiently big) such that $\|D\sigma_0(t)\|\leq \omega_0, 
 \forall t\in[t_0+\overline{\delta},+\infty)$. 
\end{lem}
\begin{pf}
First, let the consensus error $\xi=D\sigma_0$. Then,
$$
\dot{\xi}(t) = Dd(t) - D\Lambda_0\sgn{D^T\xi(t)}{\frac{m}{m+1}}.
$$
Choose the Lyapunov function candidate $V(\xi)=(1/2)\xi^T\xi$. Hence, in the interval starting from $\|\xi(t_0)\|\geq \omega_0$ and in which $\|\xi(t)\|\geq \omega_0$ is maintained, the following is satisfied
\begin{equation*}
    \begin{aligned}
&\dot{V}=\xi^T\dot{\xi} = \xi^T\left(Dd - D\Lambda_0\sgn{D^T\xi}{\frac{m}{m+1}} \right)\\
&\leq -\overline{\lambda}_0 \xi^TD\sgn{D^T\xi}{\frac{m}{m+1}} + \|\xi\|\|Dd\|\\
&\leq -\overline{\lambda}_0\|D^T\xi\|^{\frac{m}{m+1}+1} + s_{\max}(D)\|\xi\|\|d\| \\
&\leq -\omega_0\left(\overline{\lambda}_0(s_{\min}(D)\omega_0)^{\frac{m}{m+1}+1}  - s_{\max}(D)\overline{d}\right) \leq -\eta
\end{aligned}
\end{equation*}
where Corollary \ref{cor:alpha_norm}-(\ref{cor:alpha_norm_upper}) and Proposition \ref{prop:induced_norm} were used, and by choosing $\overline{\lambda}_0\geq
(s_{\min}(D)\omega_0)^{\frac{-m}{m+1}-1}(\eta \omega_0^{-1}+s_{\max}(D)\overline{d})$ for any $\eta>0$. Henceforth, $V$ will decay towards the origin with rate $\eta$ until the condition $\|\xi\|\geq \omega_0$ is no longer maintained. Hence, in order to reach such condition before the interval $[t_0,t_0+\overline{\delta}]$ ends, choose $\eta\geq \overline{\delta}^{-1}(V(\xi(t_0)) - (1/2)\omega_0^2)$ for any $\overline{\delta}>0$. Therefore, by the comparison Lemma \cite[Lemma 3.4]{khalil}, $\dot{V}\leq -\eta$ implies 
\begin{equation*}
\begin{aligned}
V(\xi(t)) &\leq V(\xi(t_0)) - \eta(t-t_0) \\&  \leq V(\xi(t_0))-\eta\overline{\delta}\leq (1/2)\omega_0^2 
\end{aligned}
\end{equation*}
for $t\in[t_0,t_0+T]$ where $t_0+T\leq t_0+\overline{\delta}$ is the moment in which $\|\xi(t_0+T)\|=\omega_0$. 
Then, the condition $\|\xi(t)\|=\|D\sigma_0(t)\|\leq \omega_0$ will be reached and maintained $\forall t\in[t_0+\overline{\delta},+\infty)$ concluding the proof. 
\end{pf}
We also show that $\dot{\sigma}_0(t)$ can be driven towards an arbitrarily small vicinity of the origin. Hence, $\dot{\sigma}_0$ can play the role of $\theta$ in the results from Appendix \ref{sec:levant}. 
\begin{lem}
\label{lem:tv_fo}
Let $\mathcal{G}$ be a tree, consider system \eqref{eq:fo_sigma0} under the same conditions from Lemma \ref{lem:fo_contraction} and the additional condition that there exists $\tilde{d}>0$ such that $\|\dot{d}(t)\|\leq \tilde{d}, \forall t\in[t_0,+\infty)$. Then, for any $\tilde{\delta}>0$ and any $\tilde{\omega}_0>0$ there exists $0<\tilde{\lambda}_0= \min \text{ \normalfont diag }\Lambda_0$ (sufficiently big) such that $\|\dot{\sigma}_0(t)\|\leq \tilde{\omega}_0, \forall t\in[t_0+\tilde{\delta},+\infty)$. 
\end{lem}
\begin{pf}
Let $\gamma_i$ be the $i$-th component of $D^TD\sigma_0$. Then, by the fact that $\frac{\text{\scriptsize d}}{\text{\scriptsize d} t}\sgn{\gamma_i}{\frac{m}{m+1}} = \frac{m}{m+1}|\gamma_i|^{\frac{m}{m+1}-1}\dot{\gamma}_i $, hence,
\begin{equation*}
\begin{aligned}
\frac{\dd}{\dd t}\sgn{D^TD\sigma_0}{\frac{m}{m+1}} &= \frac{m}{m+1}\begin{bmatrix}
|\gamma_1|^{\frac{m}{m+1}-1}\dot{\gamma}_1 \\
\vdots\\
|\gamma_\ell|^{\frac{m}{m+1}-1}\dot{\gamma}_\ell
\end{bmatrix}\\&=
\frac{m}{m+1}J(t)D^TD\dot{\sigma}_0
\end{aligned}
\end{equation*}
where $J(t) = \text{diag}\left(\left[|\gamma_1|^{\frac{m}{m+1}-1},\dots, |\gamma_\ell|^{\frac{m}{m+1}-1}\right]\right)$. Now, let change of variables $\zeta=D\dot{\sigma}
_0$ which leads to
$$
\dot{\zeta}(t)=D\dot{d}(t) -\frac{m}{m+1}D\Lambda_0J(t)D^T\zeta(t) 
$$
Additionally let $\overline{\lambda}_0\leq\tilde{\lambda}_0$ with $\overline{\lambda}_0$ chosen such that $\|D^TD\sigma_0(t)\|\leq \omega_0, \forall t\in[t_0+\overline{\delta},+\infty)$ from Lemma \ref{lem:fo_contraction}. Then, each component $\gamma_i$ will satisfy $|\gamma_i|^{\frac{m}{m+1}-1}\geq \omega_0^{\frac{m}{m+1}-1}, \forall t\in[t_0+\overline{\delta},+\infty)$ since $\frac{m}{m+1}-1<0$.
Choose the Lyapunov function $V(\zeta)=(1/2)\zeta^T\zeta$ and an arbitrary $Z>0$. Hence, in the interval starting from $\|\zeta(t_0)\|\geq Z$ and in which $\|\zeta(t)\|\geq Z$ is maintained, the following is satisfied
\begin{equation*}
\begin{aligned}
&\dot{V}=\zeta^T\left(D\dot{d} -\frac{m}{m+1}D\Lambda_0JD^T\zeta \right)\\
&\leq s_{\max}(D)\|\zeta\|\|\dot{d}\|-\frac{\tilde{\lambda}_0m}{m+1}\zeta^TDJD^T\zeta \\
&\leq  s_{\max}(D) \tilde{d}\|\zeta\|-\frac{\tilde{\lambda}_0m}{m+1}c(\mathcal{G})\omega_0^{\frac{m}{m+1}-1}\|\zeta\|^2\\& \leq
-\|\zeta\|\left(\frac{\tilde{\lambda}_0m}{m+1}c(\mathcal{G})\omega_0^{\frac{m}{m+1}-1}\|\zeta\| - s_{\max}(D) \tilde{d}\right)\\&\leq
-Z\left(\frac{\tilde{\lambda}_0m}{m+1}c(\mathcal{G})\omega_0^{\frac{m}{m+1}-1}Z- s_{\max}(D) \tilde{d}\right) \leq -\eta
\end{aligned}
\end{equation*}
by the fact that $\zeta^TDJD^T\zeta = \omega_0^{\frac{m}{m+1}-1}\zeta DD^T\zeta \geq \omega_0^{\frac{m}{m+1}-1}c(\mathcal{G})\|\zeta\|^2$ using Proposition \ref{prop:graph}-(\ref{prop:graph_connectivity}), with $c(\mathcal{G})$ as the algebraic connectivity of $\mathcal{G}$ and by choosing
$$
\tilde{\lambda}_0\geq\max\left\{\overline{\lambda}_0
,\frac{m+1}{mZc(\mathcal{G})}\omega_0^{1-\frac{m}{m+1}}\left(s_{\max}(D)\tilde{d} + Z^{-1}\eta\right)\right\}
$$
for any $\eta>0$. From this point, the proof follows exactly as the proof of Lemma \ref{lem:fo_contraction} to conclude that $\|\zeta\|\leq Z$ will be reached an maintained for $t\in[t_0+\tilde{\delta},+\infty)$ for any $Z,\tilde{\delta}>0$. Hence, since $\mathcal{G}$ is a tree, we can choose $Z = \tilde{\omega}_0s_{\min}(D)$ and obtain $\|\dot{\sigma}_0\|\leq s_{\min}(D)^{-1}\|D\dot{\sigma}_{0}\|\leq s_{\min}(D)^{-1}\|\zeta\|\leq \tilde{\omega}_0$, which concludes the proof.
\end{pf}
Using these results, we provide the proof of the contraction property for tree graphs.
\begin{lem}
\label{lem:contraction}
Consider \eqref{eq:main_algo_vec_error} and $\mathcal{G}$ to be a tree. Then, for any $0<\omega_\mu<\Omega_\mu$, $T>0$, there exists some gain matrices $K_0,\dots,K_m$ (with sufficiently big diagonal entries) such that any trajectory of \eqref{eq:main_algo_vec_error} satisfying $\|\sigma_\mu(t_0)\|\leq \Omega_\mu$ will satisfy  $\|\sigma_\mu(t)\|\leq \omega_\mu, \forall t\in[T,+\infty), \forall\mu\in\{0,\dots,m\}$.
\end{lem}
\begin{pf}
Let $\dot{\sigma}_\mu=[\dot{\sigma}_{\mu,1},\dots,\dot{\sigma}_{\mu,n}]^T$. Since $|\dot{\sigma}_{\mu,i}|\leq\|\dot{\sigma}_\mu\|$ then, from Lemma \ref{lem:integral_bounded} we know that for any $\delta$ there exists $K>0$ such that $\int_{t_0}^{t_0+\delta}|\dot{\sigma}_{0,i}|\dd t\leq \int_{t_0}^{t_0+\delta}\|\dot{\sigma}_{0}\|\dd t\leq K$. Henceforth, from Propositions \ref{prop:levant_boundedness} and \ref{prop:levant_contraction} in Appendix \ref{sec:levant} we can choose an arbitrary $\Omega_\mu'$ with $\Psi_\mu>\Omega_\mu'>\Omega_\mu$ such that there exists $\delta>0$ for which $\|\sigma_{\mu}\|\leq \Omega'_\mu\leq \Psi_\mu, \forall t\in[t_0,t_0+\delta]$. Moreover, $
\|\dot{\sigma}_1\|\leq \|\sigma_2\|+s_{\max}(K_1)\ell^{\frac{1}{2m+2}}\|D^TD\sigma_0\|^{\frac{m-1}{m+1}}\leq \Psi_2 +s_{\max}(K_1)\ell^{\frac{1}{2m+2}}(s_{\max}(D^TD)\Psi_1)^{\frac{m-1}{m+1}}$. Therefore, both $\|\sigma_1\|$ and $\|\dot{\sigma}_1\|$ remain bounded in the interval $t\in[t_0,t_0+\delta)$ and will remain bounded (by the same bounding constants) in $t\in[t_0+\delta,+\infty)$ by Proposition \ref{prop:levant_contraction} provided that $\|\dot{\sigma}_0(t)\|\leq\tilde{\omega}_0, \forall t\in[t_0+\delta,+\infty)$ and sufficiently small $\tilde{\omega}_0$. Choose $0<\tilde{\delta}<\delta$. Hence, we can identify $d(t)=\sigma_{1}$ from  Lemma \ref{lem:fo_contraction} and Lemma \ref{lem:tv_fo} and choose $\tilde{\lambda}_0 = \min\text{ diag }\Lambda_0 = \min\text{ diag }K_0$ big enough such to obtain $\|\dot{\sigma}_0(t)\|\leq\tilde{\omega}_0, \forall t\in[t_0+\delta,+\infty)$ and $\|D{\sigma}_0(t)\|\leq {\omega}_0, \forall t\in[t_0+\delta,+\infty)$, obtaining the contraction for $\|\sigma_0\|$. Contraction for $\|\sigma_\mu\|, \mu>0$ follows directly from Proposition \ref{prop:levant_contraction}, since $\tilde{U}^{(m+1)}\in[-\tilde{L},\tilde{L}]^\ell$, and by adjusting $\tilde{\omega}_0<\overline{\theta}$, concluding the proof .
\end{pf}
\subsection{Contraction for general connected graphs}
Now, in order to show the same contraction property but for general graphs, consider the following setting. Let $\mathcal{G}_A$ and $\mathcal{G}_B$ be two graphs with corresponding incidence matrices $D_A = [\tilde{D}_A,D_s]$ and $D_B = [\tilde{D}_B,D_s]$ where $D_s$ corresponds to the edges which appear in both $\mathcal{G}_A$ and $\mathcal{G}_B$. Suppose that protocol \eqref{eq:main_algo_vec2} works for each of the graphs. Then, we aim to conclude that the protocol works for their union by means of switching between them, and applying the averaging principle. However, in average, the edges that appear in both graphs contribute twice to the protocol. 

Hence, we take advantage of the different gains per-edge to attenuate such contribution. This is, choose some gain matrices $K_\mu^A$ and $K_\mu^B, \forall \mu\in\{0,\dots,m\}$ for each graph to implement protocol \eqref{eq:main_algo_vec2} in the following form: $K_\mu^A=\text{blockdiag}(2\tilde{K}_\mu^A,K_\mu^s)$ and $K_\mu^B=\text{blockdiag}(2\tilde{K}_\mu^B,K_\mu^s)$ where $K_\mu^s$ corresponds to gains for the edges in $D_s$ and $2\tilde{K}_\mu^A, 2\tilde{K}_\mu^B$ for $\tilde{D}_A,\tilde{D}_B$ respectively. Now, to study the switching between $\mathcal{G}_A$ and $\mathcal{G}_B$, let 
$$
\begin{aligned}
&F_\mu(t,\tilde{Y}_0;\varepsilon) =\\&\left\{ \begin{array}{ll}
D_AK_\mu^A\sgn{D_A^T\tilde{Y}_{0}}{\frac{m-\mu}{m+1}}, & t-t_0\in[0,\varepsilon/2)\\
D_BK_\mu^B\sgn{D_B^T\tilde{Y}_{0}}{\frac{m-\mu}{m+1}}, & t-t_0\in[\varepsilon/2,\varepsilon)\\
F_\mu(t-\varepsilon,\tilde{Y}_0;\varepsilon), & t-t_0\geq \varepsilon
\end{array}\right. 
\end{aligned}
$$
and write the dynamics of $\tilde{Y}_\mu$ for this switching protocol as a differential inclusion,
\begin{equation}
\begin{aligned}
\label{eq:main_algo_vec_proj}
    \dot{\tilde{Y}}_{\mu} &= \tilde{Y}_{\mu+1} - F_\mu(t,\tilde{Y}_0;\varepsilon), \text{\ for } 0\leq\mu \leq m-1 \\
    \dot{\tilde{Y}}_{m} &\in [-L,L]^n - F_m(t,\tilde{Y}_0;\varepsilon) \\
\end{aligned}
\end{equation}
since $PU^{(m+1)}\in[-L,L]^n$ by Assumption \ref{as:bounded}. 

Note that explicit dependence of time in \eqref{eq:main_algo_vec_proj} comes only from the terms of the form $F_\mu(t,\tilde{Y}_0;\varepsilon)$ and therefore from switching. Now, we obtain the average system, by averaging the right hand side of \eqref{eq:main_algo_vec_proj} in the interval $t-t_0\in[0,\varepsilon)$. Note that terms of the form $F_\mu(t,\tilde{Y}_0;\varepsilon)$ are averaged as
$$
\begin{aligned}
&\frac{1}{\varepsilon}\int_{t_0}^{t_0+\varepsilon} F_\mu(t,\tilde{Y}_0;\varepsilon)\dd t\\
&=\frac{1}{2}\left(D_AK_\mu^A\sgn{D_A^T\tilde{Y}_{0}}{\frac{m-\mu}{m+1}}+D_BK_\mu^B\sgn{D_B^T\tilde{Y}_{0}}{\frac{m-\mu}{m+1}}\right)\\
&=D_{AB}K_\mu^{AB}\sgn{D_{AB}^T\tilde{Y}_{0}}{\frac{m-\mu}{m+1}}
\end{aligned}
$$
where $D_{AB} = [\tilde{D}_A,\tilde{D}_B,{D_s}]$ is the incidence matrix of the superposition of $\mathcal{G}_A$ and $\mathcal{G}_B$, and $K_{\mu}^{AB} = \text{blockdiag}(\tilde{K}_\mu^{A},\tilde{K}_\mu^{B},{K}_\mu^{s})$. From, this we obtain the following conclusion about \eqref{eq:main_algo_vec_proj} with respect to the averaged version of it.
\begin{lem}
\label{lem:average}
Let $\tilde{Y}_\mu(t), 0\leq \mu\leq m$ be a solution of \eqref{eq:main_algo_vec_proj} with initial conditions $\tilde{Y}_\mu(t_0)$ and let the averaged system  
\begin{equation}
\begin{aligned}
\label{eq:main_algo_vec_mean}
    \dot{\tilde{Y}}^a_{\mu}(t) &= \tilde{Y}^a_{\mu+1}(t) + D_{AB}K_\mu^{AB}\sgn{D_{AB}^T\tilde{Y}^a_{0}(t)}{\frac{m-\mu}{m+1}}\\ &\text{for } 0\leq\mu \leq m-1, \\
    \dot{\tilde{Y}}^a_{m}(t) &\in [-L,L]^n + D_{AB}K_m^{AB}\sgn{D_{AB}^T\tilde{Y}^a_0(t)}{0} \\
\end{aligned}
\end{equation}
with $\tilde{Y}^a_\mu(t_0)=\tilde{Y}_\mu(t_0)$. Then, for any $r>0$, there exists $R>0,\varepsilon>0$ such that
$$
\|\tilde{Y}_\mu(t) - \tilde{Y}^a_\mu(t)\|\leq r, \ \ \ \forall t\in[t_0,t_0+R/\varepsilon)
$$
\end{lem}
\begin{pf}
Note that the right hand sides of \eqref{eq:main_algo_vec_proj} and \eqref{eq:main_algo_vec_mean} are locally Lipschitz. Following from \cite{cortes2008}, every locally Lipschitz function at a point is one-sided Lipshitz in a neighborhood of such point. Hence, rigorous justification of the averaging argument comes from the Bogoliubov's first theorem for one-sided Lipschitz differential inclusions \cite[Section 2.2]{gamma2014}.
\end{pf}
Using this result, we show contraction for general graphs.
\begin{lem}
\label{lem:contraction2}
Consider \eqref{eq:main_algo_vec_error} and $\mathcal{G}$ to an arbitrary connected graph. Then, for any $0<\omega_\mu<\Omega_\mu$, $T>0$, there exists some gain matrices $K_0,\dots,K_m$ (sufficiently big) such that any trajectory of \eqref{eq:main_algo_vec_error} satisfying $\|\tilde{Y}_\mu(t_0)\|\leq \Omega_\mu$ will satisfy  $\|\tilde{Y}_\mu(t)\|\leq \omega_\mu, \forall t\in[T,+\infty), \forall \mu\in\{0,\dots,m\}$.
\end{lem}
\begin{pf}
In order to show the result for graphs $\mathcal{G}$ let $N$ be the dimension of its flow space and proceed by induction. The induction base with $N=0$ is shown in Lemma \ref{lem:contraction}. Now, assume that the result is true for graphs with flow space of dimension $N-1$ with contraction neighborhood of radius $\omega_\mu(N-1)$ and $\mathcal{G}_{AB}$ be any graph with flow space dimension $N$. Then, by Proposition \ref{prop:graph_decomposition} there exists two connected graphs $\mathcal{G}_A$ and $\mathcal{G}_B$ with flow space dimension $N-1$ whose union corresponds to $\mathcal{G}_{AB}$. Choose $\omega_\mu({N-1}) = \omega_\mu(N) - r$ with arbitrary $0<r<\omega_\mu(N)$ such that there exists $K_\mu
^{A}$ and $K_\mu^{B}$ for $\mathcal{G}_A$ and $\mathcal{G}_B$ respectively and $\|\tilde{Y}_\mu(t)\| \leq \omega_\mu({N-1}), \forall t\geq T$ with $T\leq \varepsilon/2$ for each of the two networks by the assumption about the $N-1$ case. Hence, since both schemes contract to an arbitrarily small neighborhood of the origin before the switching instants at $t-t_0=\varepsilon/2$ and $t-t_0=\varepsilon$, the same conclusion applies for the switching system \eqref{eq:main_algo_vec_proj} before $t-t_0=\varepsilon$. Contraction for $\mathcal{G}_{AB}$ comes from Lemma \ref{lem:average} since \eqref{eq:main_algo_vec_mean} corresponds to the dynamics in \eqref{eq:main_algo_error2} for such graph, and the bound $\|\tilde{Y}_\mu^a(t)\|\leq \omega_\mu
({N-1}) + r=\omega_\mu(N)$.
\end{pf}
\section{Parameter design for EDCHO}\label{sec:parameters}
By inspecting the results from the previous sections, in particular the proof of Lemma \ref{lem:contraction}, it can be noticed that the parameters needed for \eqref{eq:main_algo} to reach consensus are closely related to the parameters used for a Levant's differentiator to converge. In fact, all parameters, except for $k_0$ can be found using this reasoning, simplifying the parameter design methodology. This is shown in the following Corollary:
\begin{cor}
\label{lemma:parameters}
Let $\lambda_1,\dots,\lambda_m$ be parameters chosen such that \eqref{eq:levant_recursive} is finite time stable for $\theta=0$. Thus, there exists $k_0>0$ large enough such that the conclusion of Lemma \ref{lem:contraction2} follows with $k_{\mu}=\lambda_{\mu}k_{\mu-1}^{\frac{m-\mu}{m-(\mu-1)}}$ for $\mu=1,\dots,m$.
\end{cor}
\begin{pf}
First, consider the case of tree graphs. The proof follows directly from the fact that \eqref{eq:main_algo_vec_error} can be written recursively as \eqref{eq:recursive_2}. The result is then a consequence of the reasoning in Section \ref{sec:tree}, where the last $m$ equations of \eqref{eq:recursive_2} correspond precisely to a vector form of \eqref{eq:levant_recursive}. This leaves only the condition that $k_0>0$ needs to be large enough. The case of general graphs is no different, since the gains used in such scheme can be chosen the same as the ones for tree graphs, as long as the contraction time is small enough from the arguments of the proof of Lemma \ref{lem:contraction2}. However, increasing $k_0$ decreases such contraction time too, which concludes the proof.
\end{pf}
Note that finding feasible sequences of parameters $\lambda_1,\dots,\lambda_n$ for \eqref{eq:levant_recursive} is by now a well studied topic in the literature. In fact, not only in the original work  \cite{levant2003} some feasible parameters were found by computer simulation for $L=1$, but also other works such as \cite{Cruz2019} give different possible values by means of a Lyapunov function condition. Hence, these parameters can be consulted and used directly, scaled appropriately for any $L>0$.
Moreover, motivated by the methodology in \cite{levant2003}, $k_0$ can be found by computer simulation for a concrete topology, by incrementally searching for an appropriate $k_0>0$ until convergence is obtained.
\section{Convergence of EDCHO}
\label{sec:convergence}
In this section, we show the proof of Theorem \ref{th:stability}. The proof follows by  contraction and by noticing that trajectories of \eqref{eq:main_algo_vec2} are invariant to a particular transformation, referred as the homogeneity in \cite{levant2003}.  
\begin{lem}
\label{lem:homo}
Let $\eta>0$. Then, the trajectories of \eqref{eq:main_algo_error2} are preserved by the transformation
$
(t,\tilde{Y}_\mu)\mapsto(\eta t,\eta^{m-(\mu-1)}  \tilde{Y}_\mu)
$.
\end{lem}
\begin{pf}
Let $t'=\eta t$ and  $\tilde{Y}_\mu'(t')=\eta^{m-(\mu-1)}\tilde{Y}_\mu(t'/\eta)$. Then, for $\mu=0,\dots,m-1$,
\begin{equation*}
\begin{aligned}
&\frac{\dd \tilde{Y}_\mu'}{\dd t'} =  \eta^{m-(\mu-1)}\frac{\dd \tilde{Y}_\mu}{\dd t'}=\eta^{m-\mu}\dot{\tilde{Y}}_\mu \\
&= \eta^{m-\mu}\left(\eta^{-(m-\mu)}\tilde{Y}'_{\mu+1} -DK_\mu \sgn{D^T\eta^{-(m+1)}\tilde{Y}'_{0}}{\frac{m-\mu}{m+1}}\right) \\&= \tilde{Y}'_{\mu+1} -DK_\mu \sgn{D^T\tilde{Y}'_{0}}{\frac{m-\mu}{m+1}}
\end{aligned}
\end{equation*} 
Similarly, for $\mu=m$ it is obtained $\frac{\text{\scriptsize d} \tilde{Y}_m'}{\text{\scriptsize d} t'} \in [-L,L]^n- DK_m\sgn{D^T\tilde{Y}_0'}{0}$.
Then, trajectories $\tilde{Y}_\mu'(t')$ are equivalent to $\tilde{Y}_\mu(t)$. 
\end{pf}
Similarly as the work in \cite{levant2003}, both contraction and homogeneity of \eqref{eq:main_algo_vec_error} can be used to produce sequential contractions towards an equilibrium point, reaching it in a finite amount of time. 

\begin{pf}(Of Theorem \ref{th:stability})
Let \eqref{eq:main_algo_vec2} with $K_\mu=k_\mu I$ recovering \eqref{eq:main_algo}. Then, Lemma \ref{lem:contraction2} implies that for sufficiently large $k_0,\dots,k_m>0$ there exists a finite time $T_c>0$ such that if $\|\tilde{Y}_\mu(t_0)\|\leq \Omega_\mu$ then $\|\tilde{Y}_\mu(t_0+T_c)\|< \kappa_\mu \Omega_\mu$ for any $0\leq \kappa_\mu < 1$. Then, from Lemma \ref{lem:homo} the similar contraction follows: if   $\|\tilde{Y}_\mu(t_0)\|\leq\eta^{m-\mu+1}\Omega_\mu$ then $\|\tilde{Y}_\mu(t_0 + \eta T_c)\|\leq\kappa_\mu\eta^{m-\mu+1} \Omega_\mu$.  Hence, for $0<\eta<1$ choose $\kappa_\mu = \eta^{m-\mu+1}$. Therefore, convergence is shown in a sequence of countable steps for $\nu=0,1,\dots$ of sequential contraction. For $\nu=0$, let $\|\tilde{Y}_\mu(t_0)\|\leq \Omega_\mu$ be contracted to $\|\tilde{Y}_\mu(t_0+T_c)\|\leq \eta^{m-\mu+1}\Omega_\mu$. Then, for $\nu=1$, $\|\tilde{Y}_\mu(t_0+T_c)\|\leq \eta^{m-\mu+1}$ is contracted to $\|\tilde{Y}_\mu(t_0+T_c + \eta T_c)\|\leq  \eta^{2(m-\mu+1)}\Omega_\mu$. Furthermore, for any $\nu$, the contraction 
$$
\left\|\tilde{Y}_\mu(t_0+T_c(1+\eta+\eta^2+\cdots+\eta^\nu))\right\|\leq  \eta^{(\nu+1)(m-\mu+1)}\Omega_\mu
$$ 
is obtained. Hence, 
$
\lim_{\nu\to+\infty} \eta^{(\nu+1)(m-\mu+1)} = 0
$ 
and 
$$
\lim_{\nu\to+\infty}T_c(1+\eta+\eta^2+\cdots+\eta^\nu)=\frac{T_c}{1-\eta}
$$ using the geometric series. Therefore, with $T=T_c/(1-\eta)$, $\|\tilde{Y}_\mu(t)\|=0,\forall t\in[t_0+T,+\infty)$. Moreover, convergence towards the EDC property follows from the conclusion of Corollary \ref{th:steady_state}. Finally, using the parameter design from Corollary \ref{lemma:parameters} concludes the proof.
\end{pf}
\section{Simulation examples}
\label{sec:simulations}
\begin{figure}
    \centering
    \includegraphics[width=0.35\textwidth]{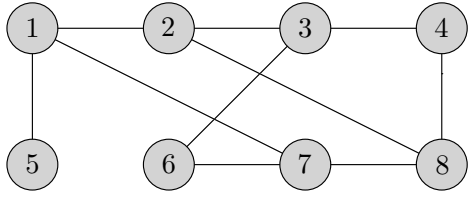}
    \caption{The graph $\mathcal{G}$ considered in the examples}
    \label{fig:graph}
\end{figure}

For the purpose of demonstrating the advantages of the proposal, a simulation scenario is described here with the following configuration. There are $n=8$ agents connected by a graph $\mathcal{G}$ shown in Figure \ref{fig:graph}. In this example we use $m=3$ and the gains $k_\mu$ are chosen as $7.5, 19.25, 17.75, 7$ for all agents. Moreover, consider initial conditions $x_{i,\mu}(0) = 0,\ \forall \mu>0$ and $x_{i,0}(0)$ given by $[18.69$, $-4.17$, $-2.02$, $-1.49$, $-4.65$, $-4.52$, $0.16$, $-2.00]$ respectively. Note that this initial conditions comply with Assumption \ref{as:initial}.  

\begin{figure}
    \centering
    \includegraphics[width=0.47\textwidth]{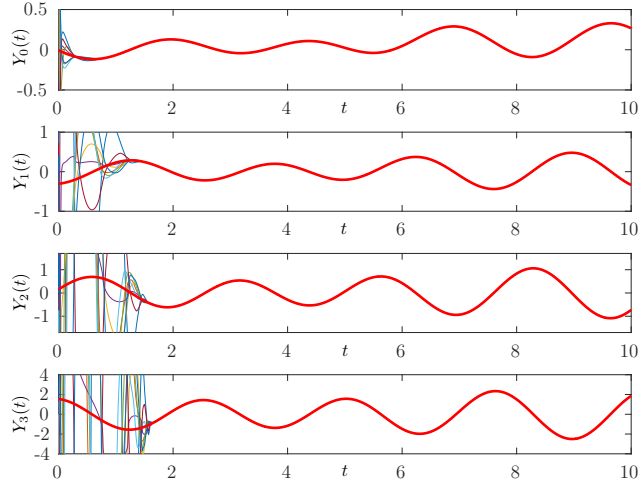}
    \caption{Components of the vectors $Y_0(t),\dots,Y_3(t)$ as well as $\bar{u}(t),\dot{\bar{u}}(t), \ddot{\bar{u}}(t)$ and ${\bar{u}}^{(3)}(t)$ in red.}
    \label{fig:traj_sine}
\end{figure}

\begin{figure}
    \centering
    \includegraphics[width=0.47\textwidth]{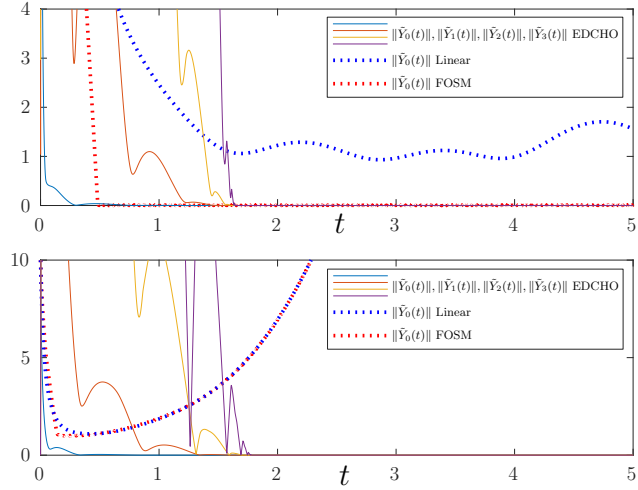}
    \caption{Comparison of the magnitude of $\tilde{Y}_\mu(t)$ for EDCHO with a similar measure for a linear protocol and a first order sliding mode (FOSM) in the case of (Up) sinusoidal and (Bottom) polynomial references signals.}
    \label{fig:traj_poly}
\end{figure}

In the first experiment, each agent has internal reference signals $u_i(t)=a_i\cos(\omega_i t+\phi_i), i=1,\dots,8$ with amplitudes $a_i$ of $[0.99$, $0.27$, $0.02$, $0.48$, $0.18$, $0.24$, $0.65$, $0.50]$, frequencies $\omega_i$ of $[2.44$, $1.70$, $1.12$, $0.26$, $0.68$, $1.73$, $2.33$, $0.02]$ and phases $\phi_i$ of $[1.25$, $1.92$, $6.25$, $3.82$, $0.70$, $7.63$, $9.93$, $6.80]$. The individual trajectories for this experiment are shown in Figure \ref{fig:traj_sine}, as well as the target $\bar{u}(t)$ and its derivatives in red. Note that all agents are able to track not only $\bar{u}(t)$ but also $\dot{\bar{u}}(t),\ddot{\bar{u}}(t)$ and $\bar{u}^{(3)}(t)$. The magnitudes of $\tilde{Y}_0,\dots,\tilde{Y}_3$ are shown in \ref{fig:traj_poly}-a), where it can be noted that exact convergence is achieved. Compare this with the behaviour of a linear protocol \cite[Equation (11)]{Solmaz2017} and the first order sliding mode (FOSM) protocol in \cite[Equation (5)]{freeman2019} under the same conditions. As shown in Figure \ref{fig:traj_poly}-(Up), the linear protocol achieves only bounded steady state error whereas the FOSM achieves exact convergence in the input and its derivatives. However, both linear and FOSM approaches are only able to track $\bar{u}(t)$ and not its derivatives. 

Now, consider the reference signals as $u_i(t)=a_i t^3, i=1,\dots,8$ with $a_i$ of $[0.99$, $0.27$, $0.02$, $0.48$, $0.18$, $0.24$, $0.65$, $0.50]$. The error convergence is shown in Figure \ref{fig:traj_poly}-(Bottom) where EDCHO achieves exact convergence whereas both the linear and the FOSM protocol errors grow to infinity.

\begin{figure}
    \centering
    \includegraphics[width=0.47\textwidth]{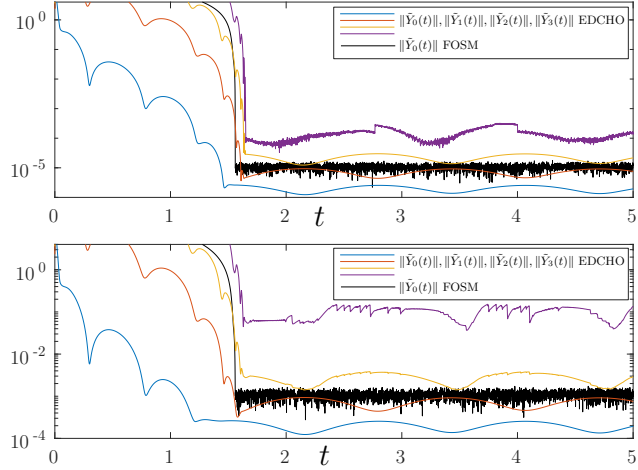}
    \caption{Comparison of the magnitude of $\tilde{Y}_\mu(t)$ for EDCHO with a similar measure for the first order sliding mode (FOSM) in the case of sampling steps of $\Delta t = 10^{-6}$ (Up) and $\Delta t = 10^{-3}$ (Bottom) sinusoidal references signals. To improve clarity, this figure is plotted in logarithmic vertical axis}
    \label{fig:h}
\end{figure}

All previous experiments were conducted by simulating the algorithms using explicit Euler's discretization with small step size of $\Delta t=10^{-6}$ in order to inspect their performance as close as possible to their continuous time theoretical version. Increasing the value of $\Delta t$ have an effect on the performance of both EDCHO and FOSM protocols due to discretization errors and chattering. Thus, we repeated the experiment with sinusoidal reference signals using $\Delta t = 10^{-3}$ instead, in order to make this effects more apparent. The results of this experiment are shown in Figure \ref{fig:h} for both $\Delta t = 10^{-6}$ and $\Delta t = 10^{-3}$. The parameters of the FOSM were chosen such to roughly match the settling time of EDCHO for the sake of fairness. Note that in both cases the error signal $\|\tilde{Y}_0(t)\|$ for EDCHO is almost one order of magnitude less than the FOSM protocol. Moreover, it can be noted that the chattering effect is almost negligible for EDCHO when compared to FOSM which greatly suffers from it. Additionally, the accuracy in steady state degrades as $\Delta t$ is increased, specially for the higher order signals $\|\tilde{Y}_\mu(t)\|, \mu>0$ as expected from HOSM systems \cite[Theorem 7]{levant2003}.
\section{Conclusions}
In this work, the EDCHO algorithm has been presented, where the agents are able to maintain zero steady-state consensus error, when tracking the average of time-varying signals and its derivatives. EDCHO works under reasonable assumptions about the initial conditions and bounds of certain high order derivatives of the reference signals. An in depth study on its stability characteristics was provided from which a simple design procedure arises. The simulation scenario presented here, exposes the effectiveness of our approach in addition to show its advantages when compared to other approaches. Nevertheless, a more general tuning procedure is yet to be explored. Similarly, we reserve the discussion on noise robustness, robustness against connection and disconnection of agents and chattering to a future work.
\appendix
\section{Some useful inequalities}
\label{sec:app_ineq}
\begin{prop}\label{prop:lp_norm} Let $x=[x_1,\dots,x_n]^T\in\mathbb{R}^n$ and define $\|x\|_p = \left(\sum_{i=1}^n|x_i|^{p}\right)^{1/p}$. Then, with $0<r<s$, the following inequalities are satisfied:
\begin{enumerate}
    \item\label{prop:lp_norm_rs} \cite[Theorem 16, Page 26]{hardy} $\|x\|_r\leq n^{\frac{1}{r}-\frac{1}{s}}\|x\|_s$.
    \item\label{prop:lp_norm_sr} \cite[Theorem 19, Page 28]{hardy} $\|x\|_s\leq \|x\|_r$.
\end{enumerate}
\end{prop}
\begin{cor}\label{cor:alpha_norm} Let $x\in\mathbb{R}^n$ and $0<\alpha<1$. Then,
\begin{enumerate}
    \item\label{cor:alpha_norm_lower} $\|\sgn{x}{\alpha}\|\leq n^{\frac{1-\alpha}{2}}\|x\|^\alpha$
    \item\label{cor:alpha_norm_upper} $x^T\sgn{x}{\alpha}\geq \|x\|^{\alpha+1}$
\end{enumerate}
\end{cor}
\begin{pf}
For the first item, note that $\|\sgn{x}{\alpha}\|^{\frac{1}{\alpha}} = \left(\sum_{i=1}^n|x_i|^{2\alpha}\right)^{\frac{1}{2\alpha}} = \|x\|_{2\alpha}$. Moreover, Proposition \ref{prop:lp_norm}-(\ref{prop:lp_norm_rs}) with $2\alpha<2$ leads to $\|x\|_{2\alpha}\leq n^{\frac{1}{2\alpha}-\frac{1}{2}}\|x\|_2=n^{\frac{1-\alpha}{2\alpha}}\|x\|$. For the second item note that $x^T\sgn{x}{\alpha} = \sum_{i=1}^n x_i\sgn{x_i}{\alpha}=\sum_{i=1}^n |x_i|^{\alpha+1}=\left(\|x\|_{\alpha+1}\right)^{\alpha+1}$. Moreover, Proposition \ref{prop:lp_norm}-(\ref{prop:lp_norm_sr}) with $\alpha+1<2$ leads to $\|x\|_{\alpha+1}\geq\|x\|_2$. Hence, $x^T\sgn{x}{\alpha}=\left(\|x\|_{\alpha+1}\right)^{\alpha+1}\geq \|x\|^{\alpha+1}$.
\end{pf}
\begin{prop}\label{prop:induced_norm}
Let $x\in\mathbb{R}$ and $A\in\mathbb{R}^{m\times n}$. Then,
$
s_{\min}(A)\|x\| \leq \|Ax\| \leq s_{\max}(A)\|x\|
$.
\end{prop}
\begin{pf}
This proposition is a direct consequence of the Rayleigh inequality \cite[Theorem 4.2.2]{horn} and the definition of the singular values of $A$ in \cite[Page 151]{horn}.
\end{pf}
\section{Auxiliary results in algebraic graph theory}
\label{sec:app_graph}
An undirected graph $\mathcal{G}=(\mathcal{V},\mathcal{E})$ consists of a node set $\mathcal{V}$ of $n$ nodes and an edge set $\mathcal{E}$ of $\ell$ edges  \cite[Page 1]{Godsil}. An edge from node $i$ to node $j$ is denoted as $(i,j)$, which means that node $i$ can communicate to node $j$ in a bidirectional way. 
$\mathcal{G}$ is said to be connected if there is a path between any two nodes. A subgraph is a cycle if every node in it has exactly two neighbors. $\mathcal{G}$ is said to be a tree, if it is connected and it has no cycles. A spanning tree is a subgraph of $\mathcal{G}$ if it contains all its nodes and is a tree. If $\mathcal{G}$ is connected, there is always a spanning tree \cite[Page 4]{Godsil}. Moreover, a tree has exactly $\ell=n-1$ edges \cite[Page 53]{Godsil}. Furthermore, we define the union of two undirected graphs $\mathcal{G}_A=(\mathcal{V}_A,\mathcal{E}_A)$ and $\mathcal{G}_B=(\mathcal{V}_B,\mathcal{E}_B)$ as $\mathcal{G}_{AB}=(\mathcal{V}_A\cup\mathcal{V}_B,\mathcal{E}_A\cup\mathcal{E}_B)$.
\begin{defn}[Matrices of interest for $\mathcal{G}$] The following matrices are defined for $\mathcal{G}$:
\begin{enumerate}
    \item \cite[Page 163]{Godsil} The adjacency matrix $A\in\mathbb{R}^{n\times n}$ of an undirected graph is defined by its components $[A]_{ij}$ which comply $a_{ij} = [A]_{ij}\triangleq 1$ if $(i,j)\in\mathcal{E}$ and $a_{ij}\triangleq 0$ otherwise.

    \item \cite[Page 167]{Godsil} An incidence matrix $D\in\mathbb{R}^{n\times \ell}$ for $\mathcal{G}$ has a column per edge, where all elements of the column corresponding to edge $(i,j)$ are $0$ except for the $i$-th element which is 1 and the $j$-th which is $-1$.
    
    \item  \cite[Page 279]{Godsil} The Laplacian matrix of $\mathcal{G}$ is defined as $Q\triangleq DD^T\in\mathbb{R}^{n\times n}$.
    
    \item \cite[Page 305]{Godsil} For any connected graph $\mathcal{G}$, the algebraic connectivity $c(\mathcal{G})>0$ is defined as the second smallest eigenvalue of $Q$.
\end{enumerate}
\end{defn}
\begin{prop}[Some algebraic properties of $\mathcal{G}$]\label{prop:graph}
Let $\mathcal{G}$ be a connected undirected graph, $x\in\mathbb{R}^n$ be any vector orthogonal to $\mathds{1}$ and $N$ be the dimension of the null space of $D$ (flow space). Then,
\begin{enumerate}
    \item\label{prop:graph_rank}  \cite[Lemma 13.1.1]{Godsil} $\text{rank}(Q)=n-1$ .
    \item\label{prop:graph_connectivity} \cite[Corollary 13.4.2]{Godsil} $x^TQx\geq c(\mathcal{G})x^Tx$.
    \item\label{prop:graph_ones} \cite[Page 280]{Godsil} $D^T \mathds{1}= 0$.
    \item\label{prop:graph_flow} \cite[Theorem 14.2.1]{Godsil} $N=\ell-n+1$.
\end{enumerate}
\end{prop}
\begin{prop}
\label{prop:graph_decomposition}
Let $\mathcal{G}_{AB}=(\mathcal{V},\mathcal{E})$ be an undirected connected graph with flow space of dimension $N>0$. Then, there exists undirected connected graphs $\mathcal{G}_A,\mathcal{G}_B$ over the same nodes $\mathcal{V}$, with flow space of dimension $N-1$ whose union is $\mathcal{G}_{AB}$. 
\end{prop}
\begin{pf}
Choose any spanning tree $\mathcal{G}_{AB}^{\text{\tiny tree}}$ of $\mathcal{G}_{AB}$ with $\ell_{\text{\tiny tree}} = n-1$ edges. Then, from Proposition \ref{prop:graph}-(\ref{prop:graph_flow}), $\ell-\ell_{\text{\tiny tree}} = N>0$. Consider first $N=1$. Hence, there is exactly one edge which isn't part of the spanning tree. Moreover, denote this edge as $e=(i,j)$ where $i,j\in\mathcal{V}$. Then, since $e$ is not in the spanning tree, it is in a cycle and $i,j$ have at least two neighbors each. Therefore, there are at least two ways to reach $i$ and $j$ from other nodes. Consequently there are at least two different spanning trees $\mathcal{G}_A,\mathcal{G}_B$ (with flow space of dimension $N-1=0$) which contain $e$. Now, for $N\geq2$, there exists at least two edges $e,e'$ which are not in the spanning tree. Let $\mathcal{G}_A$ and $\mathcal{G}_B$ be $\mathcal{G}_{AB}$ without $e$ and $e'$ respectively. These graphs are connected over the same node set $\mathcal{V}$ since they contain the same spanning tree of $\mathcal{G}_{AB}$. Moreover, they have flow space of dimension $N-1$ by Proposition \ref{prop:graph}-(\ref{prop:graph_flow}) since they have one edge less than $\mathcal{G}_{AB}$.
\end{pf}
\begin{prop}\label{prop:vector_decomposition} Let $\mathcal{G}$ be connected. Then any $x\in\mathbb{R}^n$ can be written as $x=\alpha\mathds{1}+D\tilde{x}$ with $\alpha\in\mathbb{R}$ and $\tilde{x}\in\mathbb{R}^\ell$.
\end{prop}
\begin{pf}
Let $\lambda_1,\dots,\lambda_n$ and $v_1,\dots,v_2$ be the eigenvalues and eigenvectors of $Q$ respectively with $\|v_i\|=1, i=1,\dots,n$.
First, from Propositions \ref{prop:graph}-(\ref{prop:graph_rank}) and \ref{prop:graph}-(\ref{prop:graph_ones}) we know that $Q$ has $\lambda_1=0$ and that $(1/\sqrt{n})\mathds{1}$ is its only eigenvector. Hence, $\{(1/\sqrt{n})\mathds{1},v_2,\dots,v_n\}$ is an orthonormal basis of $\mathbb{R}^n$ by the spectral Theorem \cite[Theorem 2.5.6]{horn}. Consequently, any vector $x\in\mathbb{R}^n$ can be decomposed as a vector in the image of $Q$ and a component parallel to $\mathds{1}$, equivalently $x=\alpha\mathds{1}+Qy$ for  $\alpha\in\mathbb{R}$ and $y=\mathbb{R}^n$. Additionally, let $\tilde{x}=D^Ty$ obtaining $x=\alpha\mathds{1}+DD^Ty=\alpha\mathds{1}+D\tilde{x}$. 
\end{pf}
\section{Auxiliary results on exact differentiation}\label{sec:levant}
In this section we provide some results that were used in \cite{levant2003} to show the stability of the Levant's arbitrary order exact differentiator. In particular, we are interested in the properties of the recursive system
\begin{equation}
\label{eq:levant_recursive}
    \begin{aligned}
    \dot{\sigma}_1(t) &= \sigma_{2}(t) - \lambda_1\sgn{\sigma_1(t)+\theta(t)}{\frac{m-1}{m}}\\
    \dot{\sigma}_\mu(t) &= \sigma_{\mu+1}(t) - \lambda_\mu\sgn{\sigma_\mu(t)-\dot{\sigma}_{\mu-1}(t)}{\frac{m-\mu}{m-(\mu-1)}}\\& \text{\normalfont for } 1<\mu<m \\
    \dot{\sigma}_m(t)&\in-\lambda_m\sgn{\sigma_m(t)-\dot{\sigma}_{m-1}(t)}{0}+[-L,L]
    \end{aligned}
\end{equation}
with $\sigma_\mu\in\mathbb{R}, 1\leq \mu\leq m$, and the measurable map $\theta:\mathbb{R}_+\to[-\bar{\theta},\bar{\theta}]$ with $\bar{\theta}>0$. Two important results regarding the contraction property of \eqref{eq:levant_recursive} are given.
\begin{prop}[Arbitrary boundedness of \eqref{eq:levant_recursive}]\cite[Lemma 7]{levant2003}\label{prop:levant_boundedness}
Let $\theta:\mathbb{R}_+\to[-\bar{\theta},\bar{\theta}]$ satisfy the condition  that $\int_{t_0}^{t_0+\delta}|\theta(\tau)|\dd\tau<K$ for some $K>0$. Then, for any $0<\Omega_\mu<\Omega_\mu', 0\leq \mu\leq m$ there exists $\delta>0$ (sufficiently small) such that any trajectory of \eqref{eq:levant_recursive} satisfying $|\sigma_\mu(t_0)|\leq \Omega_\mu$ will satisfy $|\sigma_\mu(t)|\leq \Omega_\mu', \forall t\in[t_0,t_0+\delta]$. 
\end{prop}
\begin{prop}[Contraction property of \eqref{eq:levant_recursive}]\cite[Lemma 8]{levant2003}\label{prop:levant_contraction}
For any $0<\omega_\mu<\Omega_\mu, 1\leq\mu\leq m$ there exists $\Omega_\mu<\Psi_\mu$, $T>0$, some gains $\lambda_1,\dots,\lambda_m>0$ (sufficiently big) and $\bar{\theta}>0$ (sufficiently small) such that any trajectory of \eqref{eq:levant_recursive} satisfying $|\sigma_\mu(t_0)|\leq \Omega_\mu$ will satisfy $|\sigma_\mu(t)|\leq \Psi_\mu, \forall t\in[t_0,t_0+T]$ and $|\sigma_\mu(t)|\leq \omega_\mu, \forall t\in[T,+\infty)$. 
\end{prop}
\bibliographystyle{plain}

\end{document}